\documentclass[twocolumn,twoside,slac_two]{revtex4}
\usepackage{graphicx}
\usepackage{fancyhdr}
\usepackage{hyperref}
\newcommand{ \be }{\begin{equation}}
\newcommand{ \ee }{\end{equation}}
\newcommand{ \bea }{\begin{eqnarray}}
\newcommand{ \eea }{\end{eqnarray}}
\newcommand{ \la }{\langle}
\newcommand{ \ra }{\rangle}

\begin{document}

\title{The Glasma and the Hard Ridge}

\author{George Moschelli$^a$ and Sean Gavin,$^a$}
\affiliation{
a) Department of Physics and Astronomy, Wayne State University, 666
W Hancock, Detroit, MI, 48202, USA}

\date{\today}
\begin{abstract}
Correlation measurements indicate that excess two particle correlations extend over causally disconnected rapidity ranges. Although, this enhancement is broad in relative rapidity $\eta=\eta_1 - \eta_2$, it is focused in a narrow region in relative azimuthal angle $\phi=\phi_1 - \phi_2$. The resulting structure looks like a ridge centered at $\eta = \phi=0$.
Similar ridge structures are observed in correlations of particles associated with a jet trigger (the hard ridge) and in correlations without a trigger (the soft ridge).
The long range rapidity behavior requires that the correlation originates in the earliest stage of the collision, and probes properties of the production mechanism.
Glasma initial conditions as predicted by the theory of Color Glass Condensate and provide a and early stage correlation that naturally extends far in rapidity. We have previously shown that the soft ridge is a consequence of particles forming from an initial Glasma phase that experience a later stage transverse flow. We extend this work  to study the ridge dependance on the $p_t$ of the correlated pairs. We then determine the soft contribution to the hard ridge.
\end{abstract}


\maketitle

\maketitle

\thispagestyle{fancy}
%
%
\section{Introduction}
Correlation measurements at RHIC show an enhancement as a function of relative rapidity $\eta=\eta_1 - \eta_2$, and relative azimuthal angle $\phi=\phi_1 - \phi_2$. Centered at $\eta=\phi=0$, the ridge gains its title because the measured structure is broad in $\eta$ and narrow in $\phi$. Both triggered and un-triggered measurements show a ridge like structure, but have been considered separate phenomena.
The hard ridge measures the correlated yield of associated particles per jet trigger \cite{Putschke:2007mi}, and the soft ridge measures the number of correlated particle pairs per particle \cite{Daugherity:2008su}.
We propose a common explanation for these phenomena based on particle production in an early Glasma stage followed by radial flow \cite{Moschelli:2009tg}.

In \cite{Moschelli:2009tg} we extend the model of Ref.\cite{Gavin:2008ev}  to incorporate jet production and address the soft and hard ridges.  Long range rapidity correlations and the insight they provide on early time dynamics are our driving concerns  \cite{Dumitru:2008wn,Gavin:2008ev}. We take particle production to occur through a Glasma state, and emphasize how computed Glasma correlations can affect ridge measurements.
In Ref.\ \cite{Gavin:2008ev} we found excellent agreement with the peak amplitude and azimuthal width shown in current Au+Au data.
In the next section we extend this work here to include Cu+Cu systems. We then extend the model of Ref.\ \cite{Gavin:2008ev} to address varying $p_t$ ranges so that we may address the hard ridge. 
In Sec.\ \ref{sec:jets} we add a contribution of jets following the model of Ref.\  \cite{Shuryak:2007fu}. We extend that model to compute both the strength and azimuthal dependence of the jet contribution to the hard and soft ridges. We then combine the flow and jet effects and find that  correlations of thermally produced pairs constitute a significant contribution to the triggered measurement. We then discuss how experiments might distinguish the different contributions. 
%
%
\section{Glasma Correlations}
The theory of Color Glass Condensate (CGC) predicts an early Glasma stage in a high energy collision in which particles are produced by strong longitudinal color fields. As nuclei collide, the transverse fields of each nucleus are instantaneously transformed into longitudinal fields that are approximately uniform in rapidity. These fields are essentially random over transverse distances $r_t$ larger than the saturation scale $Q_s^{-1}$, where $Q_s\sim 1-2$~GeV. We can think of such field configurations as consisting of a collection of longitudinal flux tubes. Flux tubes are ubiquitous in QCD-based descriptions of high energy collisions. In the Glasma they are closely packed and not strictly distinct due to saturation. They are, however, uncorrelated for $r_t > Q_s^{-1}$. This is their essential feature for this work. The Glasma changes to plasma as particles form from the fields and thermalize. 

In the saturation regime, the number of gluons in a rapidity interval $\Delta y$ is
%
\be
N = ({{dN}/{dy}})\Delta y  \sim {\alpha_s}^{-1}Q_s^2R_A^2,
\label{eq:Nscale}
\ee
where $R_A$ is the nuclear radius and $\alpha_s$ is the strong coupling constant at the saturation scale $Q_s$  \cite{Kharzeev:2000ph}.  We understand (\ref{eq:Nscale}) as the number of flux tubes $ K \sim (Q_sR_A)^2$ times the density of gluons per flux tube $\propto \alpha_s^{-1}$.  
The scale of correlations is set by 
%
\be
{\cal R} = {{\langle N^2\rangle - \langle N\rangle^2 - \langle N\rangle}\over{\langle N\rangle^2}},
\label{eq:Rdef}
\ee
where the brackets denote an average over collision events.
This quantity vanishes for uncorrelated gluons, for which multiplicity fluctuations are necessarily Poissonian. In Ref.~\cite{Gavin:2008ev} we argued that the Glasma correlation strength is ${\cal R}\propto \langle K\rangle^{-1} = (Q_sR)^{-2}$, and found that the Glasma contribution to correlations is 
%
\be
{\cal R}{{dN}/{dy}} \sim {\alpha_s}(Q_s)^{-1},
\label{eq:CGCscale}
\ee
a result consistent with calculations of Dumitru et al. in Ref.~\cite{Dumitru:2008wn}. Equations (\ref{eq:Nscale}) and (\ref{eq:CGCscale}) constitute initial conditions for the hydrodynamic evolution of the system. 

These Glasma initial conditions affect the final state correlations in several ways. First, particles emitted from the same tube share a common origin that is localized to a very small transverse area, since $Q_s \ll R_A$. Second, the flux tubes correlate particles over a large pseudorapidity range. Some of the flux tubes can stretch across the full longitudinal extent of the system at times $< 1$~fm. These flux tubes rapidly fragment. At later times, the particles they produce can be separated by large longitudinal distances depending on their momenta. Consequently, subsequent scattering and hydrodynamic evolution cannot erase their correlations -- they are causally disconnected.   
Third, the strength of the correlation depends on the number of flux tubes. The number of tubes depends on the centrality and energy of the collision, as well as the transverse area of the tube, all of which, in turn, depend on $Q_s$. Finally, as a result of the common origin, particles coming from the same tube must have the same initial radial position and feel the same effects from flow, independent if their rapidity.

During the Glasma phase, flux tubes thermalize into partons and pressure builds as the systems moves toward an equilibrated Quark Gluon Plasma. Partons initially localized in tubes are now localized in small fluid cells with a uniform azimuthal distribution of particles. Following a Hubble-like expansion of the system, the transverse fluid velocity takes the form $\gamma_t\mathbf{v}_t = \lambda \mathbf{r}_t$. All of partons in a fluid cell are boosted radially depending on their initial radial position. Consequentially, partons in any given fluid cell gain transverse momentum in the radial direction and the relative angle between any two momentum vectors in that cell becomes smaller. Furthermore, fluid cells at a larger radial position have a larger final transverse velocity and the relative angle between parton momentum vectors is narrower. This angular narrowing depends only on the initial radial position and the small transverse area of the flux tube source. In our simplistic view, all flux tubes are uniform in rapidity and extend to the same longitudinal length. This provides for a longitudinally uniform fireball that experiences the same radial flow at every longitudinal position. In this way, at every rapidity, angular correlations are enhanced in the same way and the initial state spatial correlations are both preserved through freeze out and represented in momentum correlations.

Following \cite{Gavin:2008ev} we define the momentum space correlation function at freeze out as 
%
\be
r (\mathbf{p}_1, \mathbf{p}_2) = \rho_2 (\mathbf{p}_1, \mathbf{p}_2) - \rho_1(\mathbf{p}_1)\rho_1( \mathbf{p}_2) 
\label{eq:MomCorr0}
\ee
where $\rho_2 (\mathbf{p}_1, \mathbf{p}_2) = dN/dy_1d^2p_{t1}dy_2d^2p_{t2}$ is the pair distribution, and $ \rho_1(\mathbf{p}) \equiv dN/dyd^2p_t$ is the single particle spectrum. We describe the effect of flow using the familiar blast-wave model \cite{Schnedermann:1993ws,Kiyomichi:2005zz, Barannikova:2004rp, Iordanova:2007vw, Retiere:2003kf}. In this model, the single particle spectrum is $ \rho_1(\mathbf{p}) = \int f(\mathbf{x},\mathbf{p}) \, d\Gamma$,  where $f(\mathbf{x},\mathbf{p}) =  {(2\pi)^{-3}}\exp\{-p^\mu u_\mu/T\}$ is the Boltzmann phase-space density and $d\Gamma = p^\mu d\sigma_{\mu}$ is the differential element of the Cooper-Frye freeze out surface. We assume a constant proper time freeze out, so that $d\Gamma = \tau_F m_t \cosh(y-\eta)d\eta d^2 r_t$, where $\eta = (1/2)\ln((t+z)/(t-z))$ is the spatial rapidity. 

We argue in Ref.~\cite{Gavin:2008ev} that the final-state momentum space correlation function is 
 %
\be
r(\mathbf{p}_1, \mathbf{p}_2) =
\!\!\int c(\mathbf{x}_1, \mathbf{x}_2) 
\frac{f(\mathbf{x}_1,\mathbf{p}_1)}{n_1(\mathbf{x}_1)} 
\frac{f(\mathbf{x}_2,\mathbf{p}_2)}{n_1(\mathbf{x}_2)}
d\Gamma_1d\Gamma_2, 
\label{eq:MomCorr}
\ee
where $n_1(\mathbf{x}) = \int f(\mathbf{x},\mathbf{p})d\Gamma$. The spatial correlation function $c(\mathbf{x}_1, \mathbf{x}_2)$ depends on the Glasma conditions as follows
%
\be
c(\mathbf{x}_1, \mathbf{x}_2)
 = {\cal R}\,\delta(\mathbf{r}_t ) \rho_{{}_{FT}} (\mathbf{R}_t),
\label{eq:param}
\ee
where $\mathbf{r}_t = \mathbf{r}_{t1} -  \mathbf{r}_{t2}$ is the relative transverse position, and  $\mathbf{R}_t = (\mathbf{r}_{t1} +  \mathbf{r}_{t2})/2$ is the average position. The delta function accounts for the fact that Glasma correlations are highly localized to $r_t < Q_s^{-1}$. 
The factor $\rho_{_{FT}}(\mathbf{R}_t)$ describes the transverse distribution of the flux tubes in the collision volume, which we assume follows the thickness function of the colliding nuclei.

We comment that the form of (\ref{eq:param}) holds as long as $\cal R$ is unmodified from its initial Glasma value by particle production and hydrodynamic evolution. This is only strictly true as long as a) subsequent evolution doesn't change the relative number of particles in the rapidity interval of interest; and b) the number of observed hadrons is proportional to the initial number of gluons. Causality prevents these effects from altering $\cal R$ for truly long range correlations,  $|\eta_1 - \eta_2| > 1-2$. This would hold for smaller rapidities in Glasma theory as long as boost invariance is a reasonable approximation, since $dN/dy$ is then a hydrodynamic constant of motion in each event. Moreover, assumption (b) is common in Glasma/CGC calculations. On the other hand, for $|\eta_1 - \eta_2| < 1-2$,  the experimental $dN/dy$ is not flat and will change with time due to particle diffusion and number changing processes; see \cite{Gavin:2006xd}. For now, we will assume that $\cal R$ is constant and defer the hydrodynamic modification for later work.

The analysis  in Ref.~\cite{Gavin:2008ev} focused on a measurement of  the near side peak of the soft ridge in 200 GeV Au+Au using the observable $\Delta \rho/\sqrt{\rho} = (\rho_{sib} - \rho_{ref})/\sqrt{\rho_{ref}}$  \cite{Daugherity:2008su}. The quantity $\rho_{sib}(\phi , \eta)$ represents the distribution of ``sibling pairs" from the same event, as a function of relative pseudorapidity and relative azimuthal angle, and is comparable to our $\rho_2$. The quantity $\rho_{ref}(\phi , \eta)$ represents uncorrelated pairs from mixed events and is equivalent to the square of our $\rho_1$. The difference $(\rho_{sib} - \rho_{ref})$ is a measure of correlated pairs and is comparable to the integral of (\ref{eq:MomCorr}) over the transverse momenta and average azimuthal angle $\Phi=(\phi_1+\phi_2)/2$. It is convenient to compute the quantity 
%
\be
\frac{\Delta\rho}{\rho_{ref}}=
\frac{\int r(\mathbf{p}_1, \mathbf{p}_2)p_{t1} p_{t2} dp_{t1} dp_{t2} d\Phi}
{\int \rho_1(p_{t1}) \rho_1(p_{t2}) d^2p_{t1} d^2p_{t2} },
\label{eq:dratio}
\ee
which is independent of the overall scale of the multiplicity. We emphasize that this quantity includes correlated pairs in which both particles can have any momentum, while measurements of the hard ridge correlate particles from different $p_t$ ranges. We will extend our approach to address such quantities  below.

To construct the observed quantity $\Delta\rho/\sqrt{\rho_{ref}}$, we notice that $r(\mathbf{p}_1, \mathbf{p}_2)$ computed from eq. (\ref{eq:dratio}) is proportional to the correlations strength ${\cal R}$ (\ref{eq:Rdef}). We can write 
%
\be
\frac{\Delta\rho}{\rho_{ref}}={\cal R}F(\phi),
\label{eq:dratioNorm}
\ee
where $F(\phi)$ is normalized such that  $\int_{0}^{2\pi} F(\phi) d\phi =1$. The distribution $F(\phi)$ depends only on the blast-wave parameters $\gamma m/T$ and $v_s$, and represents the angular correlations of particles from flux tubes after hydrodynamic expansion. The factor ${\cal R}$ scales the strength of the correlations with both energy and centrality and determines the rapidity dependence (which is flat in this case). 

We now combine (\ref{eq:dratio}) and (\ref{eq:Nscale}) to obtain the observed quantity 
%
\be
\Delta \rho/\sqrt{\rho_{ref}} = \kappa{\cal R} dN/dy~F(\phi),
\label{eq:deltaRho}
\ee
where we equate the factor ${\cal R}dN/dy$ with (\ref{eq:CGCscale}), which accounts for all of the Glasma energy and centrality dependence. The scale constant $\kappa$ is independent of energy. 
As described in \cite{Gavin:2008ev} we set $\kappa $ only for Au+Au 200GeV collisions such that $F(\phi)$ for the most central collisions is aligned with the most central data point. Although blast-wave parameters have some energy dependence, the Glasma factor (\ref{eq:CGCscale}) allows for strong agreement with the 62~GeV data without further adjustment of $\kappa$. 
%
\begin{figure}
\centerline{\includegraphics[width=3.2in]{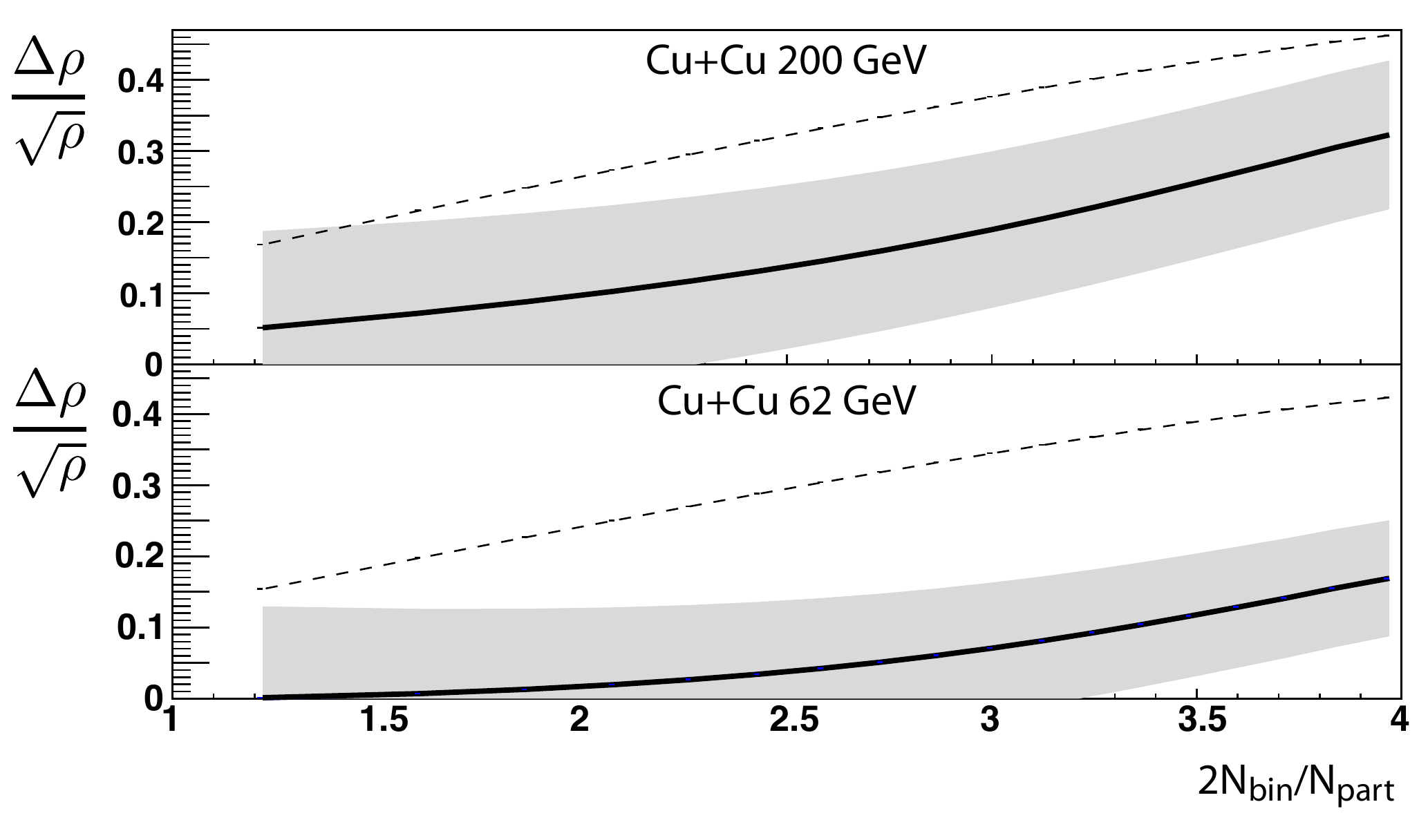}}
\caption[]{Cu+Cu amplitude as a function of centrality ($\nu=2N_{bin}/N_{part}$) for both 200 and 62 GeV. Dashed lines represent the blast wave blast-wave integrations $F(\phi)$ normalized to 200 GeV Au+Au as in \cite{Gavin:2008ev}. Solid lines represent the inclusion of the CGC scaling, ${\cal{R}}dN/dy$, following (\ref{eq:deltaRho}) with the inclusion of  (\ref{qscu}). The parameters are unchanged from Ref.\ \cite{Gavin:2008ev}}
\label{fig:fig1}\end{figure}

We now apply (\ref{eq:deltaRho}) and  (\ref{eq:CGCscale}) to address new and forthcoming data for $\Delta\rho/\sqrt{\rho_{ref}}$ and the azimuthal width Cu+Cu for both 200 and 62 GeV as a function of centrality.  No parameters are adjusted from \cite{Gavin:2008ev}  -- in that sense, these results are predictions.  The $Q_s^2$ has a dependence on the density of participants $\rho_{\rm part}$ that is determined in Ref.~ \cite{Kharzeev:2000ph}. The saturation scale in central Cu+Cu is then
%
\be
Q_s({\rm Cu})^2=Q_c({\rm Au})^2\frac{\rho_{\rm part}({\rm Cu,\, central})}{\rho_{\rm part}{(\rm Au,\, central})}.
\label{qscu}
\ee
After this scaling, we take the relative centrality dependence of $Q_s$ to be the same in the Cu and Au systems. This assumption can be refined by measuring the centrality dependence of $dN/dy$ in Cu+Cu. Similarly, we obtain the blast wave parameters $T$ and $v_s$ in Cu+Cu from the Au+Au values by assuming that they scale with the number of participants. These assumptions can be refined as single particle spectra measurements as a function of centrality become available.

Figure \ref{fig:fig1} shows our prediction for the soft ridge amplitude in Cu+Cu 200 and 62 GeV systems as a function of centrality. In both panels, the dashed line represents the blast-wave amplitudes $F(\phi=0)-F(\phi=\pi)$ normalized to fit Au+Au at 200 GeV in Ref.\ \cite{Gavin:2008ev}. The dashed lines are included to show how the energy and system-size dependence affects the calculation when the Glasma dependence (\ref{eq:CGCscale}) is omitted. The solid lines are the result of including the Glasma scaling (\ref{eq:CGCscale}) adjusted by (\ref{qscu}). The error band represents a $~10\%$ uncertainty in the blast-wave parameters plus an additional uncertainty in the parameterization of $Q_s$ that increases with decreasing centrality.

\begin{figure}
\centerline{\includegraphics[width=3.2in]{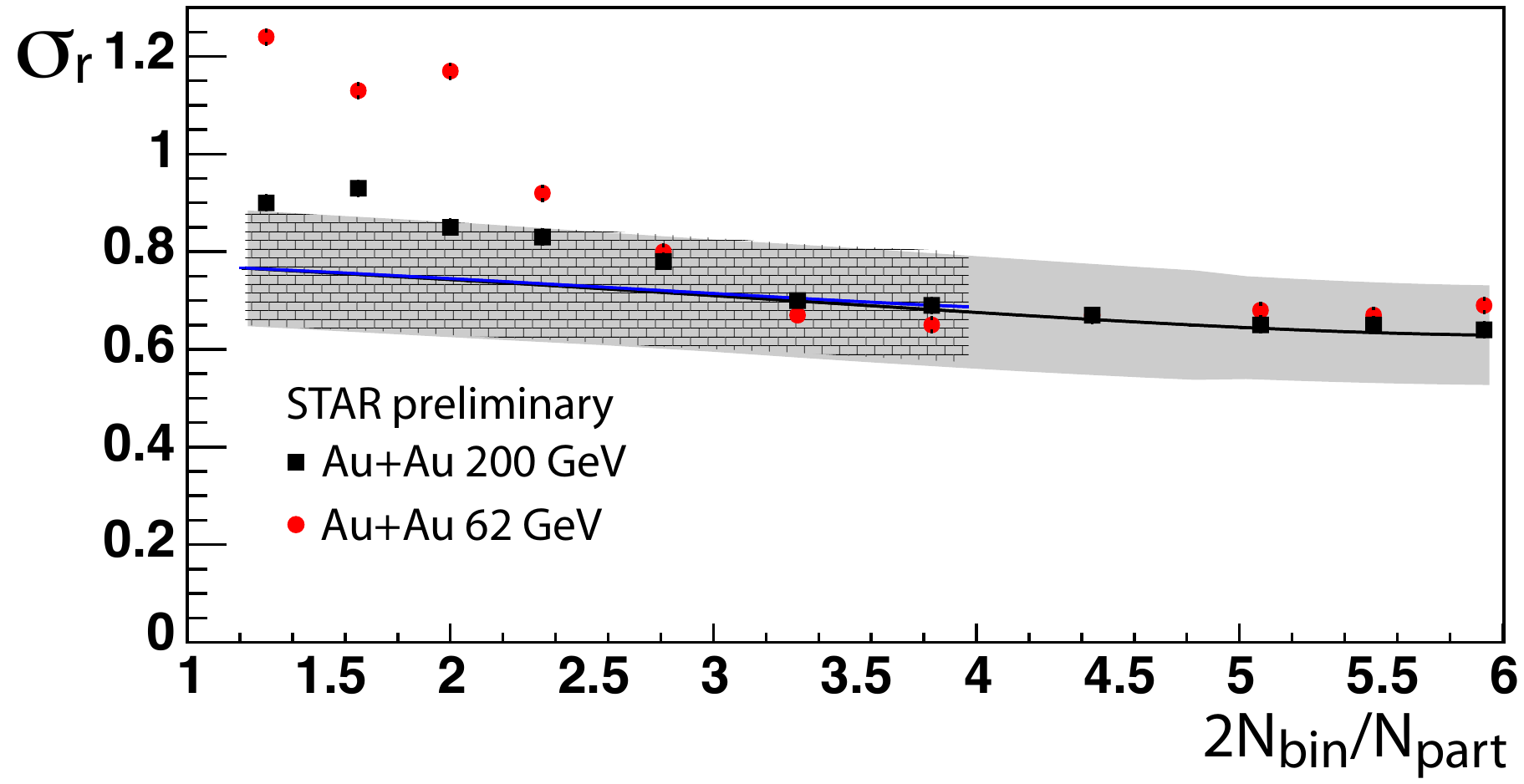}}
\caption[]{Comparison of the previously published angular width calculations for Au+Au (black line) with the angular widths for Cu+Cu systems (blue line, hatched error band) compared with preliminary STAR Au+Au 200  and 62 GeV data. Width calculations remain independent of energy and nearly independent of system}
\label{fig:fig2}\end{figure}

In Fig.\ \ref{fig:fig2} we show the soft ridge azimuthal width in Cu+Cu systems compared to previously published Au+Au result \cite{Gavin:2008ev}. We have also included preliminary STAR measurement  of the soft ridge in Au+Au width \cite{Daugherity:2008su}. In \cite{Gavin:2008ev} we find that the azimuthal width of the near side peak of the soft ridge in Au+Au is due to radial flow, is constant with a change in energy, and is relatively uniform with change in centrality. The error band is representative of the uncertainty of fitting an offset gaussian to the angular calculation. Since the azimuthal width is completely determined by radial flow, which depends completely on the choice of centrality and blast-wave parameterizations, and all of those parameters have remained unchanged, we calculate the same enhancement in the width for Cu+Cu as Au+Au. The black line extending to $\nu=6$ is the Au+Au result from \cite{Gavin:2008ev}, and the overlaid  blue line extending to $\nu=4$ with the hatched error band is the Cu+Cu result. Again, as with the Au+Au result, the Cu+Cu result is independent of energy since the measured transverse expansion does not depend on energy. 
%

\begin{figure}
\centerline{\includegraphics[width=3.2in]{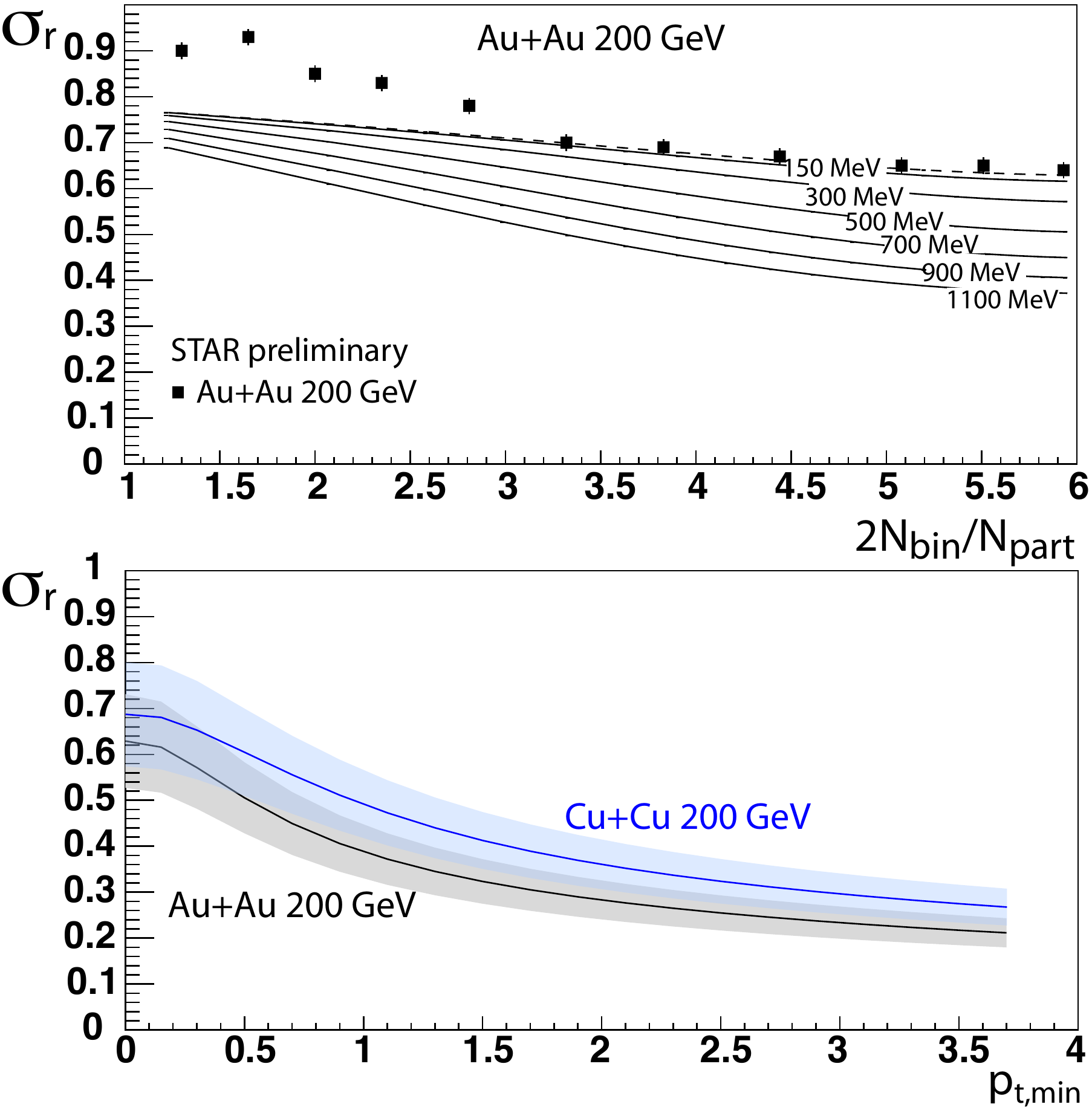}}
\caption[]{Top panel shows the angular width for Au+Au systems with increasing minimum $p_{t,min}$ limits compared with preliminary STAR Au+Au 200 GeV data. The dashed line represents the $p_{t,min}$=0 calculation shown in Fig.\ \ref{fig:fig2}. The lower panel shows the azimuthal width for most central collisions vs the $p_{t,min}$ limit for both Au+Au and Cu+Cu 200 GeV.}
\label{fig:fig2a}\end{figure}

A key feature of the flow-based descriptions of the ridge is that it is the angular width, $\sigma_r$, of correlated pairs decreases as the mean $p_t$ of the pair increases.  The greater the radial boost given to a fluid cell, the narrower the relative angle between the momentum vectors of particles in that cell. A very high momentum correlated pair is more likely to have come from a fluid cell that received a very large transverse boost. To study whether this effect is present, we compute $\Delta \rho/\sqrt{\rho_{ref}}$ for pairs of $p_t > p_{t,\,{\rm min}}$. As $p_{t,\,{\rm min}}$ increases, we also expect the amplitude of the ridge to decrease, since it is more difficult to find higher $p_t$ bulk particles. We therefore compute
%
\begin{eqnarray}
 \left( \frac{\Delta\rho}{\rho_{ref}} \right)_{p_{t}} =
\frac{
\int \limits_{p_{t2,min}}^{p_{t2,max}} \int \limits_{p_{t1,min}}^{p_{t1,max}} 
r(\mathbf{p}_1, \mathbf{p}_2)
}
{\int \limits_{p_{t1,\rm{min}}}^{p_{t1,max}} \rho_1(p_{t1})
 \int \limits_{p_{t2,min}}^{p_{t2,max}}  \rho_1(p_{t2}) 
 },\nonumber  \\
\nonumber \\
={\cal{R}}F(\phi;p_{t1,min},p_{t1,max},p_{t2,min},p_{t2,max}),
\label{eq:ptmin}
\end{eqnarray}
where the integration measures are the same as in (\ref{eq:dratio}). To obtain the measured ratio we write 
%
%
\be
\frac{(\Delta \rho/\sqrt{\rho_{ref}})_{p_t}}{\Delta \rho/\sqrt{\rho_{ref}}} = 
\frac{F(\phi;p_{t,min},\infty,p_{t,min},\infty)}{F(\phi)}
\frac{\int_{p_{t1,min}}^\infty \rho_1}{\int_0^\infty \rho_1}.
\label{eq:CGCpt}
\ee
For increasing values of $p_{t,min}$ we calculate the correlation function as before using (\ref{eq:ptmin}) and (\ref{eq:CGCpt}) and find that the azimuthal width does indeed decrease as shown in the upper panel of Fig.\ \ref{fig:fig2a}.
%
%
%

\begin{figure}
\centerline{\includegraphics[width=3.2in]{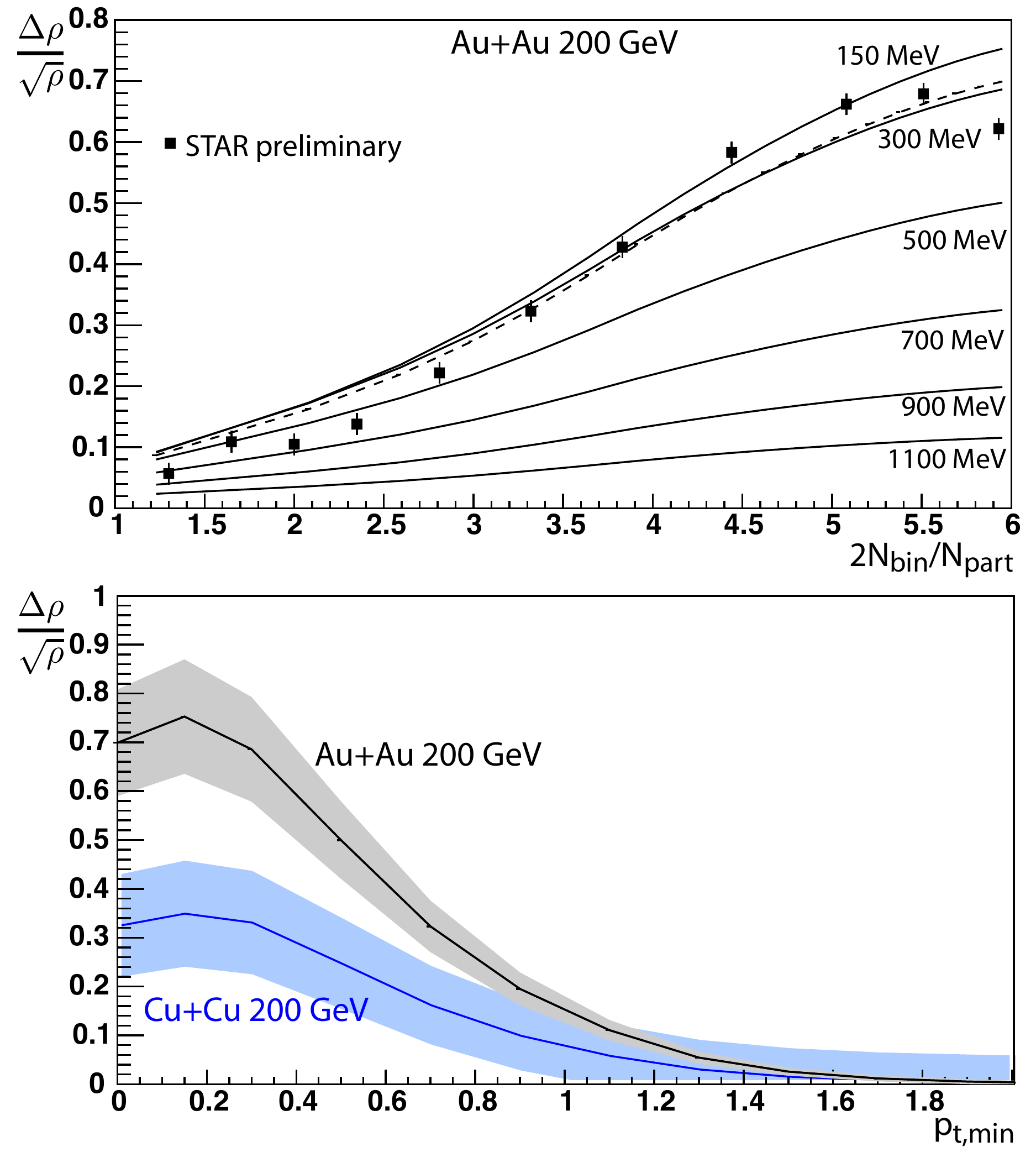}}
\caption[]{Top panel: Au+Au 200 GeV amplitude calculations for increasing $p_{t,min}$ limits. The dashed line is calculation with $p_{t,min}=0$. Lower panel: the soft ridge amplitude for most central collisions plotted as function of the $p_{t,min}$ limit for both Au+Au and Cu+Cu 200 GeV.}
\label{fig:fig3}\end{figure}

The upper panel of Fig.\ \ref{fig:fig3} shows the correlation amplitude vs. centrality for different choices of $p_{t,min}$. The amplitude decreases with increasing $p_{t,min}$ because the number of particles contributing to correlations is reduced. The lower panel of Fig.\ \ref{fig:fig3} shows the amplitude of $\Delta\rho / \sqrt{\rho_{ref}}$ for the most central collision as a function of the choice of $p_{t,min}$ for Au+Au and Cu+Cu at 200 GeV. Similarly, the blue curve in the lower panel of Fig.\ \ref{fig:fig2a} represents the azimuthal width of the soft ridge in most central collisions as a function of $p_{t,min}$.  

We see that the azimuthal width of the hard ridge is smaller than that of of the soft ridge, but as the $p_{t,min}$ limit of the soft ridge is increased, the amplitude of the correlations drops and the azimuthal width narrows. In the $p_t$ range of the hard ridge, it appears that the azimuthal width could be narrow enough, but the amplitude is not directly comparable.  To understand this difference, we must understand the differences in the two measurements. The most significant difference is the choice of the momentum range of the correlated particles. The hard ridge measurement analyzes the yield of associated particles per jet trigger where the associated particle $p_t$ range and the trigger range do not overlap. The soft ridge measurement, however, finds the number of correlated pairs per particle where both particles are in the same range with $p_t$ above minimum bias. The normalization of the soft ridge is found by taking the square root of the uncorrelated pair reference spectrum. 

STAR measures the hard ridge, or yield of associated particles per jet trigger, for Au+Au 200 GeV for $3<p_{t,trigg}<4$ GeV with $2<p_{t,assoc}<3$ \cite{Putschke:2007mi}. Identifying $p_{t1}$ with the trigger range and the associated range with $p_{t2}$, we calculate $\Delta\rho / \sqrt{\rho_{ref}}$ and transform to yield by
%
\begin{eqnarray}
{\rm Yield}=  
\left( \frac{\Delta\rho}{\sqrt{\rho_{ref}}} \right)_{p_t}
\left( \frac{\int\rho_1(p_{t2})}{\sqrt{\int\rho_1(p_{t1})\int\rho_1(p_{t2})}} \right).
\label{eq:yield}
\end{eqnarray}
At higher ranges of $p_{t1,2}$ the contribution from jets should become more significant. It is important therefore to know the relative contribution of thermal particles and jet particles. As will be discussed in more detail later,  we decompose the total particle spectrum into thermal bulk and jet fractions, and to obtain the contribution of bulk correlations to the hard ridge, we multiply (\ref{eq:yield}) by the bulk fraction $\int\rho_1(p_{t1})/\int\rho_{tot}(p_{t1})$  where $\int\rho_{tot}(p_{t1})$ is the total number of particles in the range of $p_{t1}$.
%
%
%

The blue curve in Fig.\ \ref{fig:fig4} represents the contribution to the hard ridge from only thermal bulk pairs. As can be seen on the figure, bulk-bulk correlations contribute significantly to the amplitude of the triggered measurement, but seems to have a somewhat narrow profile in azimuth. It was shown in \cite{Shuryak:2007fu} that a jet acquires angular correlations with flowing matter due to quenching, but the width of the correlation is wider than the data. The contribution of jet correlations with bulk particles could make up the difference between the blue curve in Fig.\ \ref{fig:fig4} and the data by increasing both the amplitude and the width of the calculation.

In the next section we combine a theory angular correlations from \cite{Shuryak:2007fu} with spatial correlations of jets and flux tubes to obtain a jet-bulk contribution to the hard ridge.
%
%
\section{Jets, Glasma, and Correlations}\label{sec:jets}
As the $p_t$ of correlated particles is increased, the contributions from jets should become prevalent, particularly for small $\eta$. At small rapidity differences, correlations of jet particles with fragments should be large, but restricted to the size of the jet cone, and the transfer of momentum from jet particles to bulk particles is causally limited to $\sim 1-2$ units in rapidity. The existence of correlations with jet particles at larger $\eta$ would require a correlation early in the collision that remains through the longitudinal expansion of the system and is still present at freeze out. Both the hard collisions and flux tubes are made in the initial moments of the nuclear collision. Assuming that the entire overlap region of the colliding nuclei is in the saturation regime, flux tubes would fill the collision volume and a jet formed at any transverse position would be accompanied by a flux tube at the same position.
Since the flux tube extends to large rapidities, the correlation of particles from the tube and the jet can extend to large $\eta$. Angular correlations arise since particles from the tube acquire a radial trajectory from flow as before, but the jet trajectory has a bias in the radial direction due to quenching \cite{Shuryak:2007fu}.

We construct a distribution of jets as follows. We assume that jets are produced with a hard scattering rate $f_0(\mathbf{p}_1)$ that is independent of position, multiplied by a spatial profile $P_{rod}\propto (1-r_1^2/R_A^2) $ that is roughly proportional to the density of binary collisions. The phase space density of jet particles is then 
\begin{eqnarray}
f_{J}(\mathbf{x}_1,\mathbf{p}_1)=
f_0(\mathbf{p}_1)P_{prod}(r_1)S(r_1,\phi_1), 
\label{eq:jet} 
\label{eq:sruvive0}
\end{eqnarray}
where $S$ is the survival probability of a jet due to jet quenching. In practice $f_0(\mathbf{p}_1)$ cancels in $\Delta \rho/\rho_{ref}$ so we need not specify it.  We follow Ref.~\cite{Shuryak:2007fu} and take  
%
\begin{eqnarray}
S(r_1,\phi_1)=exp(-L(r_1,\phi_1)/l_{abs}),
\label{eq:sruvive}
\end{eqnarray}
%
%
where $l_{abs}=0.25 fm$ is the jet attenuation length. The survival of the jet depends on the path it takes out of the medium
%
\be
L(r_1,\phi_1)=\sqrt{R_A^2-r_1^2\sin^2(\phi_1)}-r_1\cos(\phi_1).
\label{eq:jpath}
\ee
The path (\ref{eq:jpath}) is the distance a jet would have to travel out of a circular transverse area at an angle $\phi_1$ with respect to the radial vector pointing to its position of production $r_1$ \cite{Shuryak:2007fu}. In view of (\ref{eq:jpath}) and (\ref{eq:sruvive}), the shortest path, which is the path a jet is most likely to survive, is one that is radially outward from its position of production. Although a jet production is at a minimum on the surface, $r\approx R_A$, with little material for the jet to pass through, the survival probability is maximum in any direction (not pointing into the volume). The resulting angular correlations are weak, since only a small fraction of phase space is restricted by quenching. The largest probability of production would occur at the center where $L\approx R_A$ in all directions, but there are no correlations in this case since quenching would be maximum. More concisely, jets are less likely to be produced at the surface and have a wide angular distribution. The most jets are produced the center and would have the narrowest correlation with radially flowing particles, but have the highest probability of being quenched. As the production point of the jet moves from the center toward the surface, the probability of production decreases, the probability of survival increases, and the angular correlation with radially flowing particles widens. Integration over all possibilities determines the width of the 
correlations.
%
\begin{figure}
\centerline{\includegraphics[width=2.8in]{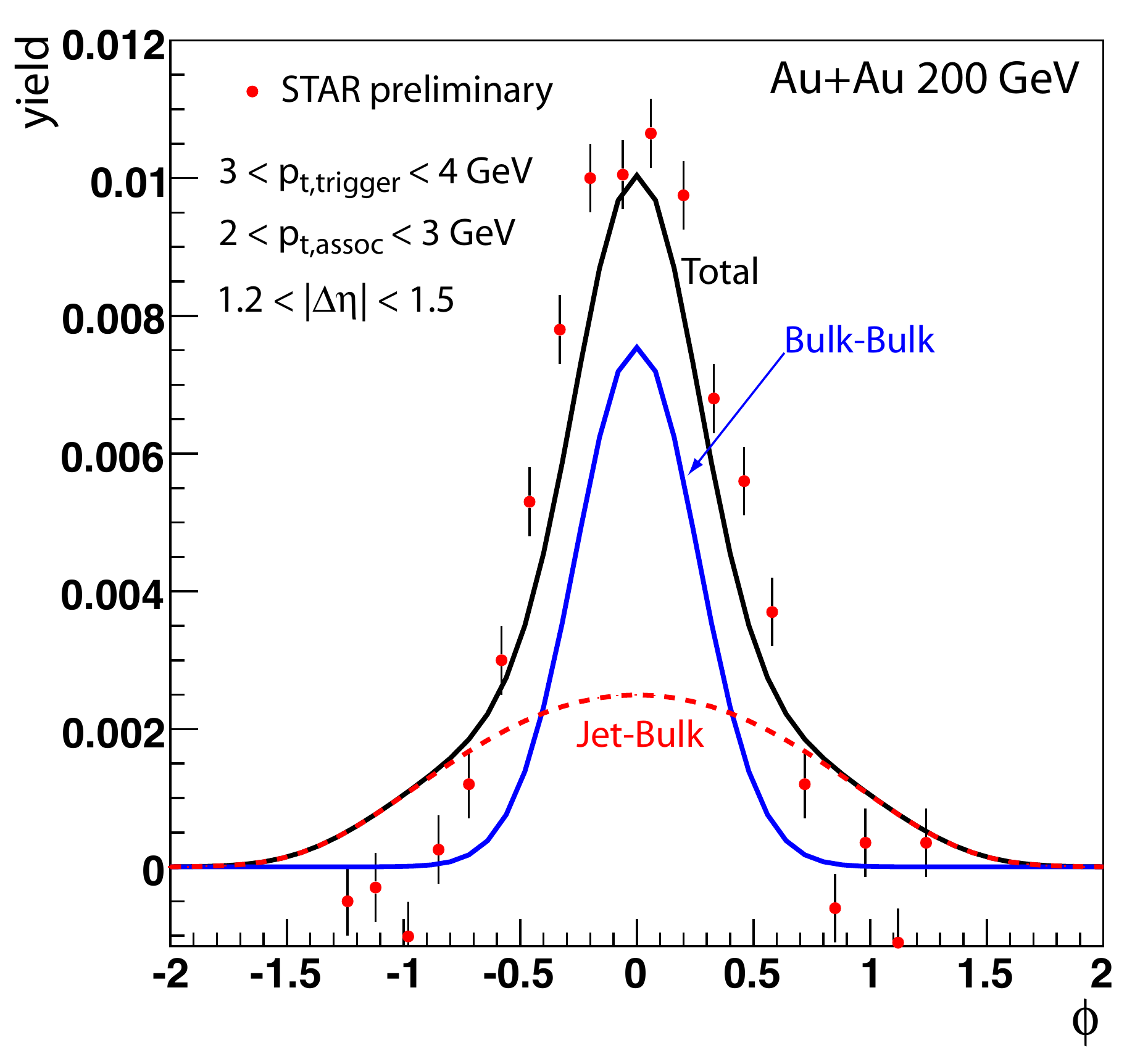}}
\caption[]{Angular profile of the jet triggered ridge in a rapidity range away from the jet peak. The solid black line combines of long range correlations. Bulk-bulk (blue line) and jet-bulk (dashed red line) contributions are shown separately. The jet fraction is determined by $p_{s}=1.25$ GeV; bulk-bulk correlations make up $\sim75\%$ of the total amplitude.}
\label{fig:fig4}
\end{figure}

The calculation follows the analysis of (\ref{eq:MomCorr}), but with the first particle from a jet and the second from a flux tube,  so that
 %
\be
r_{_{JB}}(\mathbf{p}_1, \mathbf{p}_2) =
\!\!\int c_{_{JB}}(\mathbf{x}_1, \mathbf{x}_2) 
\frac{f_{_{J}}(\mathbf{x}_1,\mathbf{p}_1)}{n_{1J}(\mathbf{x}_1)} 
\frac{f_{_{B}}(\mathbf{x}_2,\mathbf{p}_2)}{n_{1B}(\mathbf{x}_2)}
d\Gamma_1d\Gamma_2.
\label{eq:JB}
\ee
The correlation function $c_{_{JB}}(\mathbf{x}_1, \mathbf{x}_2)$ requires that a jet and a bulk particle must come from the same radial position. The rest of the equation accounts for the different the spectra for jet and bulk particles. We write
%
\be
c_{_{JB}}(\mathbf{x}_1, \mathbf{x}_2)
 =\frac{{\cal R}_{_{JB}}\la N_J\ra}{{\cal R} \la N_B\ra}
 c(\mathbf{x}_1, \mathbf{x}_2),
 \label{eq:jbcorr}
\ee
where $c$ is given by (\ref{eq:param}).

To relate the correlation strength ${\cal R}_{_{JB}}$ to the bulk correlation strength ${\cal R}$ discussed earlier, we assume that the hard scattering rate is independent of the flux tube dynamics.  Recall that the bulk quantity $\cal R$ in (2) is related to the number of flux tubes; see (\ref{eq:CGCscale}). If we take the fraction of jet and bulk particles per flux tube to be independent of the number of flux tubes, we can write  $\langle N_{_{J}}\rangle=\alpha \langle N\rangle$, where $\alpha$ and $\beta$ may depend on momentum, but do not vary event by event. We then follow  \cite{Pruneau:2002yf} to find  $\langle N_{_{B}}\rangle=\beta \langle N\rangle$, and $\langle N_{_{J}}N_{_{B}}\rangle=\alpha\beta\langle N(N-1)\rangle$, so that 
\begin{eqnarray}
{\cal R}_{_{JB}}=
\frac{\langle N_{_{J}}N_{_{B}}\rangle-\langle N_{_{J}}\rangle\langle N_{_{B}}\rangle}
{\langle N_{_{J}}\rangle\langle N_{_{B}}\rangle} \nonumber \\
=\frac{\alpha\beta\langle N(N-1)\rangle-\alpha\langle N\rangle\beta\langle N\rangle}
{\alpha\langle N\rangle\beta\langle N\rangle} \nonumber \\
=\frac{\langle N(N-1)\rangle-\langle N\rangle^2}
{\langle N\rangle^2}={\cal R}.
\label{eq:rjb}
\end{eqnarray}
Therefore, the addition of jets to the total multiplicity doesn't change the correlation strength. In essence, the beam jet associated with the hard process is just another flux tube in the high density Glasma state.

We can now rewrite (\ref{eq:deltaRho}) for jet-bulk correlations as
%
\be
\Delta \rho_{_{JB}}/\sqrt{\rho_{ref}} = \kappa{\cal R} dN_{jet}/dy\,\, F_{_{JB}}(\phi).
\label{eq:dRhoJB}
\ee
Our calculation of yield and implementation of lower $p_t$ limits follows (\ref{eq:yield}) and (\ref{eq:ptmin}) but, this time, we  scale by the jet fraction $\int\rho_{1,J}(p_{t1})/\int\rho_{1,tot}(p_{t1})$.

In order to compute the amplitude and azimuthal width of the hard ridge in Fig.~\ref{fig:fig4}, we must determine the relative contributions from both bulk-bulk and jet-bulk correlations. The measured $p_t$ spectrum follows an exponential behavior at low $p_t$ and a power law behavior where jets play a larger role, see e.g.\ Ref.\ \cite{Adams:2003kv}.  Our blast wave formulation describes the exponential behavior of the low $p_t$ spectrum well. The scale $p_s$ at which the spectrum begins to deviate from exponential behavior is proportional to $Q_s$ in Glasma theory, but the proportionality constant is not known. This introduces a free parameter -- $p_s$ at $\sqrt{s} = 200$~GeV --  that we fix below. We then find the number of jet particles by taking the difference between the total number of particles and the number of thermal particles $\rho_{1,J}=\rho_{1,tot}-\rho_{1,B}$. We take $\rho_{1,tot}$ from the measured spectrum in Ref.\  \cite{Adams:2003kv} and $\rho_{1,B}$ from the blast wave calculation with the appropriate normalization. 

We now calculate the combined effect of Glasma, flow, and jets on the correlation function. Adding the bulk-bulk and jet-bulk contributions, we obtain
%
\bea
\frac{\Delta\rho}{\sqrt{\rho_{ref}}}= 
\kappa{\cal R} \frac{dN}{dy}F_{_{BB}}(\phi)\frac{\int\rho_{1,B}(p_t)}{\int\rho_{1,tot}(p_t)} \nonumber \\ \nonumber\\
+\kappa{\cal R} \frac{dN_{jet}}{dy}F_{_{JB}}(\phi)\frac{\int\rho_{1,J}(p_t)}{\int\rho_{1,tot}(p_t)}.
\label{eq:total}
\eea
In order to compare to the yield of associated particles in the hard ridge, we combine (\ref{eq:total}) and (\ref{eq:yield}) with the appropriate integration limits.  We find that the agreement with the data in Fig.\ \ref{fig:fig4} requires $p_{s}=1.25$ GeV. The dashed red curve in Fig.\ \ref{fig:fig4} represents the contribution to the yield from correlations of jet and bulk thermal particles. This contribution is too wide. On the other hand, the bulk-bulk correlation function describing the effect of flow alone, which is  given by the blue curve, is too narrow and the computed peak height is too small. The combination of the two effects shown as the black curve in Fig.\ \ref{fig:fig4}  gives nice agreement with both the amplitude and azimuthal width. 

We now compare hard and soft ridge measurements directly by computing the momentum dependent correlation function (\ref{eq:total}).
At low $p_{t,min}$ the contribution from jets is negligible, therefore the amplitude and width of (\ref{eq:total}) is determined by the bulk-bulk term. As the $p_{t,min}$ is increased, both the amplitude and the azimuthal width of the bulk-bulk term decreases, while the amplitude of the jet-bulk term increases. 
The azimuthal width of jet-bulk correlations is roughly independent of $p_{t,min}$; the growth is due to the growth of the jet fraction. Jet-bulk correlations should become a more significant fraction of the total as $p_{t,min}$ is increased, and the azimuthal width of the ridge should increase toward the jet-bulk width.

We emphasize that the decrease of the bulk-bulk contribution to the amplitude of $\Delta\rho/\sqrt{\rho}$ and  $\sigma_r$ with increasing $p_t$ in Figs.\ \ref{fig:fig2a} and \ref{fig:fig3} is a direct consequence of transverse flow and, consequently, is a firm prediction if the jet-bulk  contribution is neglected.  The role of jets and other phenomena like recombination are less clear. We have chosen the model of Ref.\ \cite{Shuryak:2007fu} because it relies only of the well-studied phenomena of jet quenching. Our calculations using this model predict that the width would increase for higher $p_t$ ranges. 

%
%
%
\section{Summary}
We have shown that long range Glasma correlations contribute significantly to both the hard and soft ridges. Bulk-bulk correlations arise due to particle production from a Glasma state followed by transverse flow \cite{Dumitru:2008wn}, \cite{Gavin:2008ev, Moschelli:2009tg}. The Glasma initial conditions allow us to extend to study of Cu+Cu systems at both 200 and 62 GeV. We see in Fig.\ref{fig:fig1} and Fig.\ref{fig:fig2}, that we maintain good agreement with current data. To study the soft contribution to the hard ridge, we study the momentum dependence of bulk-bulk correlations and find that they are still significant, even in the $p_t$ ranges of the hard ridge measurement. To explore the effect of jet production and quenching on the ridge we included a model of jet-bulk correlations following \cite{Shuryak:2007fu}. As seen in Fig.\ref{fig:fig4} we find that we cannot explain the magnitude and width of the hard ridge without a substantial soft component.
\section*{Acknowledgments}
We thank L. McLerran, R. Bellwied, C. De Silva, A. Timmins, A. Dumitru, A. Majumder, J. Nagel, P. Sorenson, P. Steinberg, R. Venugopalan, C. Pruneau, and S. Voloshin.  This work was supported in part by U.S. NSF grants PHY-0348559 and PHY-0855369.

\bibliographystyle{G-h-physrev5}
\bibliography{references}
\vfill\eject
\end{document}